\RequirePackage{lineno}

\documentclass[aps,prl,twocolumn,superscriptaddress,showpacs]{revtex4}

\usepackage{graphicx}
\usepackage{amsmath}
\usepackage{mathtools}
\usepackage{float}
\usepackage{xspace}

\newcommand{\vmet}{\vec{{\mbox{$\protect \raisebox{0.3ex}{$\not$}E$}}}_{T}}
\newcommand{\met}{{\mbox{$\protect \raisebox{0.3ex}{$\not$}E$}}_{T}}

\begin{document}

\title{Search for a dark matter candidate produced in association with a single top quark in \boldmath{$p\bar{p}$} collisions at \boldmath{$\sqrt{s} = 1.96$} TeV}

\affiliation{Institute of Physics, Academia Sinica, Taipei, Taiwan 11529, Republic of China}
\affiliation{Argonne National Laboratory, Argonne, Illinois 60439, USA}
\affiliation{University of Athens, 157 71 Athens, Greece}
\affiliation{Institut de Fisica d'Altes Energies, ICREA, Universitat Autonoma de Barcelona, E-08193, Bellaterra (Barcelona), Spain}
\affiliation{Baylor University, Waco, Texas 76798, USA}
\affiliation{Istituto Nazionale di Fisica Nucleare Bologna, $^{ee}$University of Bologna, I-40127 Bologna, Italy}
\affiliation{University of California, Davis, Davis, California 95616, USA}
\affiliation{University of California, Los Angeles, Los Angeles, California 90024, USA}
\affiliation{Instituto de Fisica de Cantabria, CSIC-University of Cantabria, 39005 Santander, Spain}
\affiliation{Carnegie Mellon University, Pittsburgh, Pennsylvania 15213, USA}
\affiliation{Enrico Fermi Institute, University of Chicago, Chicago, Illinois 60637, USA}
\affiliation{Comenius University, 842 48 Bratislava, Slovakia; Institute of Experimental Physics, 040 01 Kosice, Slovakia}
\affiliation{Joint Institute for Nuclear Research, RU-141980 Dubna, Russia}
\affiliation{Duke University, Durham, North Carolina 27708, USA}
\affiliation{Fermi National Accelerator Laboratory, Batavia, Illinois 60510, USA}
\affiliation{University of Florida, Gainesville, Florida 32611, USA}
\affiliation{Laboratori Nazionali di Frascati, Istituto Nazionale di Fisica Nucleare, I-00044 Frascati, Italy}
\affiliation{University of Geneva, CH-1211 Geneva 4, Switzerland}
\affiliation{Glasgow University, Glasgow G12 8QQ, United Kingdom}
\affiliation{Harvard University, Cambridge, Massachusetts 02138, USA}
\affiliation{Division of High Energy Physics, Department of Physics, University of Helsinki and Helsinki Institute of Physics, FIN-00014, Helsinki, Finland}
\affiliation{University of Illinois, Urbana, Illinois 61801, USA}
\affiliation{The Johns Hopkins University, Baltimore, Maryland 21218, USA}
\affiliation{Institut f\"{u}r Experimentelle Kernphysik, Karlsruhe Institute of Technology, D-76131 Karlsruhe, Germany}
\affiliation{Center for High Energy Physics: Kyungpook National University, Daegu 702-701, Korea; Seoul National University, Seoul 151-742, Korea; Sungkyunkwan University, Suwon 440-746, Korea; Korea Institute of Science and Technology Information, Daejeon 305-806, Korea; Chonnam National University, Gwangju 500-757, Korea; Chonbuk National University, Jeonju 561-756, Korea}
\affiliation{Ernest Orlando Lawrence Berkeley National Laboratory, Berkeley, California 94720, USA}
\affiliation{University of Liverpool, Liverpool L69 7ZE, United Kingdom}
\affiliation{University College London, London WC1E 6BT, United Kingdom}
\affiliation{Centro de Investigaciones Energeticas Medioambientales y Tecnologicas, E-28040 Madrid, Spain}
\affiliation{Massachusetts Institute of Technology, Cambridge, Massachusetts 02139, USA}
\affiliation{Institute of Particle Physics: McGill University, Montr\'{e}al, Qu\'{e}bec, Canada H3A~2T8; Simon Fraser University, Burnaby, British Columbia, Canada V5A~1S6; University of Toronto, Toronto, Ontario, Canada M5S~1A7; and TRIUMF, Vancouver, British Columbia, Canada V6T~2A3}
\affiliation{University of Michigan, Ann Arbor, Michigan 48109, USA}
\affiliation{Michigan State University, East Lansing, Michigan 48824, USA}
\affiliation{Institution for Theoretical and Experimental Physics, ITEP, Moscow 117259, Russia}
\affiliation{University of New Mexico, Albuquerque, New Mexico 87131, USA}
\affiliation{The Ohio State University, Columbus, Ohio 43210, USA}
\affiliation{Okayama University, Okayama 700-8530, Japan}
\affiliation{Osaka City University, Osaka 588, Japan}
\affiliation{University of Oxford, Oxford OX1 3RH, United Kingdom}
\affiliation{Istituto Nazionale di Fisica Nucleare, Sezione di Padova-Trento, $^{ff}$University of Padova, I-35131 Padova, Italy}
\affiliation{University of Pennsylvania, Philadelphia, Pennsylvania 19104, USA}
\affiliation{Istituto Nazionale di Fisica Nucleare Pisa, $^{gg}$University of Pisa, $^{hh}$University of Siena and $^{ii}$Scuola Normale Superiore, I-56127 Pisa, Italy}
\affiliation{University of Pittsburgh, Pittsburgh, Pennsylvania 15260, USA}
\affiliation{Purdue University, West Lafayette, Indiana 47907, USA}
\affiliation{University of Rochester, Rochester, New York 14627, USA}
\affiliation{The Rockefeller University, New York, New York 10065, USA}
\affiliation{Istituto Nazionale di Fisica Nucleare, Sezione di Roma 1, $^{jj}$Sapienza Universit\`{a} di Roma, I-00185 Roma, Italy}
\affiliation{Rutgers University, Piscataway, New Jersey 08855, USA}
\affiliation{Texas A\&M University, College Station, Texas 77843, USA}
\affiliation{Istituto Nazionale di Fisica Nucleare Trieste/Udine, I-34100 Trieste, $^{kk}$University of Udine, I-33100 Udine, Italy}
\affiliation{University of Tsukuba, Tsukuba, Ibaraki 305, Japan}
\affiliation{Tufts University, Medford, Massachusetts 02155, USA}
\affiliation{University of Virginia, Charlottesville, Virginia 22906, USA}
\affiliation{Waseda University, Tokyo 169, Japan}
\affiliation{Wayne State University, Detroit, Michigan 48201, USA}
\affiliation{University of Wisconsin, Madison, Wisconsin 53706, USA}
\affiliation{Yale University, New Haven, Connecticut 06520, USA}
\affiliation{Universit\'e de Strasbourg/CNRS-IN2P3, F-67037 Strasbourg, France}

\author{T.~Aaltonen}
\affiliation{Division of High Energy Physics, Department of Physics, University of Helsinki and Helsinki Institute of Physics, FIN-00014, Helsinki, Finland}
\author{B.~\'{A}lvarez~Gonz\'{a}lez$^z$}
\affiliation{Instituto de Fisica de Cantabria, CSIC-University of Cantabria, 39005 Santander, Spain}
\author{S.~Amerio}
\affiliation{Istituto Nazionale di Fisica Nucleare, Sezione di Padova-Trento, $^{ff}$University of Padova, I-35131 Padova, Italy}
\author{D.~Amidei}
\affiliation{University of Michigan, Ann Arbor, Michigan 48109, USA}
\author{A.~Anastassov$^x$}
\affiliation{Fermi National Accelerator Laboratory, Batavia, Illinois 60510, USA}
\author{A.~Annovi}
\affiliation{Laboratori Nazionali di Frascati, Istituto Nazionale di Fisica Nucleare, I-00044 Frascati, Italy}
\author{J.~Antos}
\affiliation{Comenius University, 842 48 Bratislava, Slovakia; Institute of Experimental Physics, 040 01 Kosice, Slovakia}
\author{F.~Anz\'a}
\affiliation{Istituto Nazionale di Fisica Nucleare Pisa, $^{gg}$University of Pisa, $^{hh}$University of Siena and $^{ii}$Scuola Normale Superiore, I-56127 Pisa, Italy}
\author{G.~Apollinari}
\affiliation{Fermi National Accelerator Laboratory, Batavia, Illinois 60510, USA}
\author{J.A.~Appel}
\affiliation{Fermi National Accelerator Laboratory, Batavia, Illinois 60510, USA}
% ...

\collaboration{CDF Collaboration\footnote{With visitors from
$^a$Istituto Nazionale di Fisica Nucleare, Sezione di Cagliari, 09042 Monserrato (Cagliari), Italy,
$^b$University of CA Irvine, Irvine, CA 92697, USA,
$^c$University of CA Santa Barbara, Santa Barbara, CA 93106, USA,
$^d$University of CA Santa Cruz, Santa Cruz, CA 95064, USA,
$^e$Institute of Physics, Academy of Sciences of the Czech Republic, Czech Republic,
$^f$CERN, CH-1211 Geneva, Switzerland,
$^g$Cornell University, Ithaca, NY 14853, USA,
$^h$University of Cyprus, Nicosia CY-1678, Cyprus,
$^i$Office of Science, U.S. Department of Energy, Washington, DC 20585, USA,
$^j$University College Dublin, Dublin 4, Ireland,
$^k$ETH, 8092 Zurich, Switzerland,
$^l$University of Fukui, Fukui City, Fukui Prefecture, Japan 910-0017,
$^m$Universidad Iberoamericana, Mexico D.F., Mexico,
$^n$University of Iowa, Iowa City, IA 52242, USA,
$^o$Kinki University, Higashi-Osaka City, Japan 577-8502,
$^p$Kansas State University, Manhattan, KS 66506, USA,
$^q$Ewha Womans University, Seoul, 120-750, Korea,
$^r$University of Manchester, Manchester M13 9PL, United Kingdom,
$^s$Queen Mary, University of London, London, E1 4NS, United Kingdom,
$^t$University of Melbourne, Victoria 3010, Australia,
$^u$Muons, Inc., Batavia, IL 60510, USA,
$^v$Nagasaki Institute of Applied Science, Nagasaki, Japan,
$^w$National Research Nuclear University, Moscow, Russia,
$^x$Northwestern University, Evanston, IL 60208, USA,
$^y$University of Notre Dame, Notre Dame, IN 46556, USA,
$^z$Universidad de Oviedo, E-33007 Oviedo, Spain,
$^{aa}$CNRS-IN2P3, Paris, F-75205 France,
$^{bb}$Texas Tech University, Lubbock, TX 79609, USA,
$^{cc}$Universidad Tecnica Federico Santa Maria, 110v Valparaiso, Chile,
$^{dd}$Yarmouk University, Irbid 211-63, Jordan.
}}
\noaffiliation

\begin{abstract}
We report a new search for dark matter in a data sample of an integrated luminosity of 7.7 fb$^{-1}$ of Tevatron $p\bar{p}$ collisions at $\sqrt{s} = 1.96$ TeV, collected by the CDF II detector. We search for production of a dark matter candidate, $D$, in association with a single top quark. We consider the hadronic decay mode of the top quark exclusively, yielding a final state of three jets with missing transverse energy. The data are consistent with the standard model; we thus set 95\% confidence level upper limits on the cross section of the process $p\bar{p}\rightarrow t+D$ as a function of the mass of the dark-matter candidate. The limits are approximately 0.5 pb for a dark-matter particle with mass in the range of $0-150$ GeV/$c^{2}$.
\end{abstract}


\maketitle

Despite its successes, the standard model (SM) of particle physics leaves many important questions unanswered. For example the SM does not provide a candidate for dark matter (DM). Direct detection experiments such as DAMA\,\cite{ref:dama}, CoGeNT\,\cite{ref:cogent1,ref:cogent2} and CRESST\,\cite{ref:cresst} have reported signals suggestive of DM with mass in the few GeV/$c^2$ range, and with coupling to the SM sector of a strength enabling its detection at collider experiments. Many beyond-the-SM theories predict DM candidates to include such coupling between the DM and SM sectors.

In the framework of effective field theories, production of a DM particle ($D$) in association with a single top quark at hadron colliders has been recently studied\,\cite{ref:monotop, ref:DM, ref:models}. Here, we denote the final state containing one top quark and dark matter as {\it monotop}. Such studies are also inspired by the models of monojet produced in association with missing energy used to probe gravitons\,\cite{ref:monojet}.  Monotop DM production is described by a set of Lagrangians incorporating all possible types of DM particles (scalar, vector, fermion, {\it etc.}) and their flavor-violating interactions with quarks\,\cite{ref:IVDM, ref:IVDM2, ref:IVDM3, ref:IVDM4, ref:IVDM5}. This effective theory can inclusively describe many beyond-the-SM models. For example, new physics processes with a monotop final state can also arise from the decay of a supersymmetric squark into a neutralino and a top quark, from the decay of a vector leptoquark into a massless neutrino and a top quark, or through flavor-changing neutral interactions with a new vector state escaping detection.

In the SM, top quarks are primarily produced in pairs at particle colliders. They can also be produced singly via weak interactions, resulting in a final state consisting of a single top quark with additional lighter-flavor quarks. SM single top-quark events in the missing energy plus jets channel have been studied within the standard-model hypothesis\,\cite{ref:sper3}. As the beyond-the-SM monotop theory predicts production of a single top quark in association with a DM particle, the published SM single top-quark results do not provide any conclusive information on the existence of monotop. In addition, searches for the associated production of top quarks with DM particles have only been performed in the context of events containing a pair of top quarks\,\cite{ref:sper1, ref:sper2, Aad:2011wc}. Therefore, a dedicated search for monotops produced in colliders is needed, as the observation of monotops would be a clear sign of new physics. In this Letter, we report the first direct search for monotop signatures at particle colliders, assuming the top quark to be produced through flavor-changing interactions of up and top quarks, in association with a DM candidate $D$. We assume that the $D$ particle has a mass in the range of $0-150$ GeV/$c^2$; we do not consider decays of the $D$ particle to up and top quarks in a higher mass range.

The top quark is short-lived and decays approximately 100\% of the time into a $b$ quark and a $W$ boson, where $W \rightarrow l\nu, q\bar{q}'$. We consider the exclusive decay mode $t+D \rightarrow Wb +D$ in which $W\rightarrow q\bar{q}'$. This $W$ decay mode has the largest branching ratio and it allows for the full reconstruction of the top quark. In this channel, the missing transverse energy ($\met$)\,\cite{ref:met} can be uniquely assigned to the DM particle's passage through the detector.

Events are collected by CDF II\,\cite{ref:CDFII}, a general purpose detector used to study Tevatron $p\bar{p}$ collisions with $\sqrt{s} = 1.96$ TeV. CDF II contains a tracking system consisting of a cylindrical open-cell drift chamber and silicon microstrip detectors immersed in a 1.4 T magnetic field parallel to the beam axis. Electromagnetic and hadronic calorimeters surrounding the tracking system measure particle energies. Drift chambers and muon scintillators located outside the calorimeters identify muons. We use a data sample corresponding to an integrated luminosity of $7.7\pm0.5$ fb$^{-1}$.

We consider only those events which triggered the data acquisition system due to the presence of two calorimeter clusters and significant $\met$. We include data recorded between 2001 and 2010. Prior to 2007, the data acquisition system $\met$ threshold was 35 GeV\,\cite{ref:sper3}.  After an upgrade to the system\,\cite{ref:L2upgrade} resulting in improved jet energy and $\met$ resolution, the requirement was lowered to $\met > 30$ GeV. Jets are reconstructed using the {\sc jetclu} algorithm\,\cite{ref:jetclu} with a clustering radius of 0.4 in azimuth-pseudorapidity space ($\phi,\eta$)\,\cite{ref:pseudo}. Jet energies are corrected using standard techniques\,\cite{ref:jetcorr}. Jets originating from $b$ quarks are identified using a secondary-vertex-tagging algorithm\,\cite{ref:btag}.

In order to retain only those events for which the trigger system is fully efficient, we select events with $\met >50$ GeV and three jets. Exactly one jet is identified as a $b$-jet. We require the jet transverse energy $E_T^{j_i}$, to be $E_T^{j_1}>35$ GeV, $E_T^{j_2}>25$ GeV, $E_T^{j_{3}}>15$ GeV, where the jets $j_i$  $(i=1,2,3)$ are ordered by decreasing energy. We require that either $j_i$ or $j_2$ have $|\eta| < 0.9$, and that all three jets have $|\eta| < 2.4$. We veto events with identified high-$p_T$ electrons or muons, removing monotop events inconsistent with a hadronically-decaying top quark.

We model the signal and background contributions to the selected sample using a variety of Monte Carlo (MC) simulation programs. In our simulation we assume a top-quark mass of $172.5$ GeV$/c^2$, consistent with the world's best determination\,\cite{ref:masst, ref:topmassreview}. We model monotop DM production in the flavor-violating process ($ug\rightarrow tD$) with {\sc madgraph}\,\cite{ref:madgraph}. Additional showering and hadronization are described by {\sc pythia}\,\cite{ref:pythia}. We have generated 11 signal samples assuming various DM mass in steps of 5 GeV/$c^2$ from 0 to 25 GeV/$c^{2}$, and then in steps of 25 GeV/$c^{2}$ from 25 to 150 GeV/$c^{2}$.

The event selection described above gives a data sample dominated by QCD multijet events, where the false $\met$ arises from the mismeasurement of jet energy. Simulation of this background is prohibitive due to the high production rate and large theoretical uncertainties. Instead, we use a method which relies on data and is based on a recently improved Tag Rate Matrix (TRM) method\,\cite{ref:tagrate}. The TRM method utilizes an estimate of the probability for QCD multijet events to have tagged jets. The probability is derived in a control region dominated by QCD multijet events. This probability is applied as a per-event weight to all events meeting our analysis selections excluding the $b$-jet requirement. From this sample of weighted events, we subtract the expected electroweak components (as modeled by applying the same TRM probability to simulated samples).Ê The resulting events form our model of the QCD multijet component of the analysis data sample.

We model other physics with samples generated by MC programs. Diboson  and $t \bar{t}$ production are generated by {\sc pythia} and normalized to the next-to-leading order (NLO) cross section predicted using the {\sc mcfm} program\,\cite{ref:XsecDIB1, ref:XsecDIB2} and the approximate next-to-next-to-leading order cross section\,\cite{ref:Xsectt1}, respectively. The production of $W/Z$ plus light flavor and heavy flavor (HF) jets are simulated by {\sc alpgen}\,\cite{ref:alpgen} with showering and hadronization performed by {\sc pythia} and normalized to NLO cross sections.  Single top, both $s-$ and $t$- channel production, are modeled using {\sc madgraph} with {\sc pythia} and normalized to NLO cross sections\,\cite{ref:XsecST1, ref:XsecST2}.

The light flavor jets misidentified as $b$-jets by the secondary-vertex-tagging algorithm are labeled as {\it mistags}. A data-driven method is used to estimate the mistag rate for the tagging algorithm\,\cite{ref:btag}. We apply the mistag rate to the MC events with light flavor jets to estimate the mistag contribution.

\begin{figure}[!htb]
\includegraphics[width=0.4\textwidth]{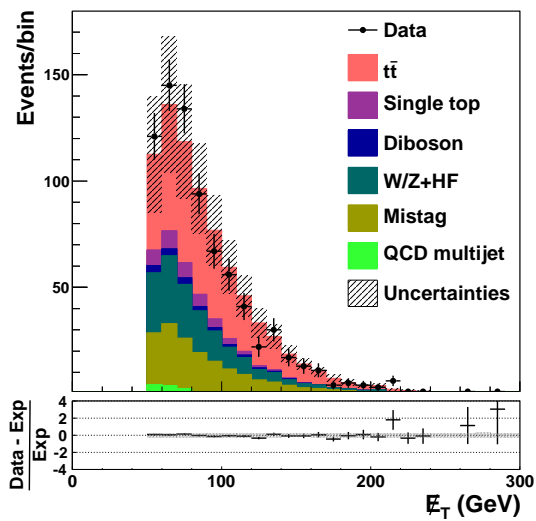}
\caption{The $\met$ distribution in a control region requiring an identified high-$p_T$ lepton; the gray area represents the statistical and systematic uncertainties on the background model. The lower panel displays the difference between the data and the expected backgrounds (Exp) divided by the latter.}
\label{ref:ewk-met}
\end{figure}

Figure\,\ref{ref:ewk-met} shows the $\met$ distribution in a control region for events which pass our signal selection but have an identified high-$p_T$ electron or muon.

After the selection described above, we are left with 6471 data events. We expect that approximately 70\% of these events come from QCD multijet production. In order to further suppress the QCD contamination and the other SM backgrounds, we require the azimuthal distances between the $\met$ and $j_2$, $\Delta \phi(\met, j_2) > 0.7$, as the $\met$ in QCD multijet background tends to align to the jet with less measured energy.  We also require the invariant mass of the three jets to be consistent with the reconstructed top-quark mass, $110 < m_{jjj} < 200$ GeV/$c^2$, large $\met$ significance ($\met/\sqrt{\sum E_T} >3.5 \sqrt{\mathrm{GeV}}$, where $\sum E_T$ is the scalar sum of transverse energy deposited in the calorimeter) and $E_T^{j_3} > 25$ GeV.  All selections have been chosen  to optimize the significance $S/\sqrt{S+B}$, where $S$ and $B$ are the expected number of signal and backgrounds events, respectively. Table \ref{ref:EAC} shows the number of events in the signal region for the data, the number of events for SM backgrounds, and the expected signal assuming different values of the DM particle's mass. The events that fail these signal-region selections are used to form a control region that is used to validate the background models, as well as to determine the normalization of the QCD multijet background.

\begin{table}[!htb] 
\begin{center}
\caption{Number of expected signal and background events compared to data in the signal region. The expected signals, assuming different values for the mass of the DM particle, are also presented. The errors include statistical and systematic uncertainties.}
\label{ref:EAC}
\begin{ruledtabular}
\begin{tabular}{cc}
Processes &  Events\\ 
\hline
$p\bar{p} \rightarrow t + D$ & \\
$m_D = 20$ GeV/$c^{2}$ & 2116.9 $\pm$ 121.4 \\
$m_D = 75$ GeV/$c^{2}$ & 232.3 $\pm$ 22.9 \\
$m_D = 100$ GeV/$c^{2}$ & 129.8 $\pm$ 12.5 \\
$m_D = 125$ GeV/$c^{2}$ & 94.5 $\pm$ 9.3 \\
\hline
$t \bar{t}$ & 182.8 $\pm$ 20.2 \\
Single top & 24.3 $\pm$ 4.5\\
Diboson & 15.7 $\pm$ 2.7\\
$W/Z$+HF & 130.5 $\pm$ 33.8\\
Mistag & 96.9 $\pm$ 39.4\\
QCD multijet & 210.2 $\pm$ 54.5\\
\hline
Total background & 660.2 $\pm$ 78.1\\
\hline
Data & 592 \\
\end{tabular}
\end{ruledtabular}
\end{center}
\end{table}

We consider several systematic uncertainties affecting the sensitivity of this search. The dominant systematic sources are the uncertainties on multijet normalization (25.5\%), the mistag rate (16.6\%) and the background cross sections ($6.5\% - 30\%$). We also consider uncertainties from the jet energy scale\,\cite{ref:jetcorr} ($2.8\% - 10.7\%$), the luminosity measurement\,\cite{ref:Lumi} (6\%), parton density functions (2\%), lepton veto (2\%), $b$-tagging efficiency (5.2\%), trigger efficiency ($0.4\% - 0.9\%$), and from the initial-state and final-state radiation (4\%). We also assign systematic uncertainties, based on the variation in the shape of the distribution of kinematic quantities, under a $\pm 1\sigma$ variation of the jet energy scale and the uncertainty on the efficiency of the data acquisition system.

\begin{figure}[!htb]
\includegraphics[scale=0.36]{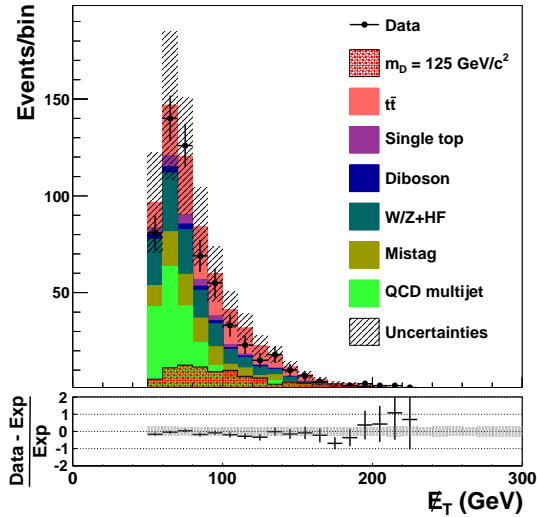}
\caption{The $\met$ distribution in the signal region. The data is compared to the sum of the SM contributions. The distribution of signal events with a DM mass of 125 GeV/$c^2$ is also shown.}\label{fig:met}
\end{figure}

\begin{figure}[!htb]
\includegraphics[scale=0.4]{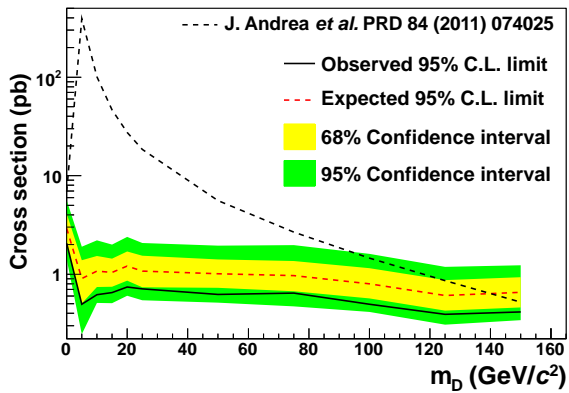}
\caption{Exclusion curve of the monotop cross section as a function of the mass of DM particle}\label{fig:final}
\end{figure}

The $\met$ is chosen to discriminate the signals from the backgrounds. The $\met$  distribution due to a DM particle of mass of 125 GeV/$c^2$ and the SM backgrounds are shown in Fig.\,\ref{fig:met}. The signal is expected to contribute significantly at high values of $\met$. We find no significant excess of signal-like events in the data analyzed, and thus proceed to set 95\% confidence level (C.L.) upper limits on the monotop DM production cross section. The limits are calculated with the $\met$ distribution as the shape discriminant using a Bayesian maximum likelihood method assuming a flat prior for the signal cross section\,\cite{ref:tom-CL}. We treat systematic uncertainties using a Bayesian marginal likelihood method. Figure\,\ref{fig:final} shows the calculated upper limits on the monotop cross section as a function of the mass of the DM candidate compared to the theoretical predictions.

In conclusion, we have performed the first search for the production of DM in association with a single top quark at hadron colliders. In an analysis of 7.7 fb$^{-1}$ of CDF II data we have found that the observed data is consistent with the expectation from SM backgrounds. We set 95\% C.L. upper limits on the cross section of $p\bar{p} \rightarrow D+t$ as a function of the DM mass in the range of $0-150$ GeV/$c^{2}$. Future searches for new physics in monotop final states can probe resonant production of top quarks and DM candidates with exotic mediators. While these processes are predicted to have low production rates (making them difficult to probe with Tevatron data), they are expected to be within the reach of LHC experiments with sufficient data.

We thank the Fermilab staff and the technical staffs of the participating institutions for their vital contributions. We also thank Fabio Maltoni for useful discussions. This work was supported by the U.S. Department of Energy and National Science Foundation; the Italian Istituto Nazionale di Fisica Nucleare; the Ministry of Education, Culture, Sports, Science and Technology of Japan; the Natural Sciences and Engineering Research Council of Canada; the National Science Council of the Republic of China; the Swiss National Science Foundation; the A.P. Sloan Foundation; the Bundesministerium f\"ur Bildung und Forschung, Germany; the Korean World Class University Program, the National Research Foundation of Korea; the Science and Technology Facilities Council and the Royal Society, UK; the Russian Foundation for Basic Research; the Ministerio de Ciencia e Innovaci\'{o}n, and Programa Consolider-Ingenio 2010, Spain; the Slovak R\&D Agency; the Academy of Finland; and the Australian Research Council (ARC).


\begin{thebibliography}{99}

\bibitem{ref:dama} R.~Bernabei {\it et al.} (DAMA Collaboration), Eur.\ Phys.\ J. C {\bf 56}, 333 (2008).
\bibitem{ref:cogent1} C.~E.~Aalseth {\it et al.} (CoGeNT Collaboration), Phys. Rev. Lett. {\bf 106}, 131301 (2011).
\bibitem{ref:cogent2} C.~E.~Aalseth {\it et al.} (CoGeNT Collaboration), Phys. Rev. Lett. {\bf 107}, 141301 (2011).
\bibitem{ref:cresst} G.~Angloher {\it et al.}, arXiv:1109.0702.
\bibitem{ref:DM} J.~Kamenik and J.~Zupan, Phys. Rev. D {\bf 84}, 111502 (2011).
\bibitem{ref:monotop} J.~Andrea, B.~Fuks, and F.~Maltoni,
 Phys. Rev. D {\bf 84}, 074025 (2011).
\bibitem{ref:models} D.~Alves {\it et al.}, arXiv: 1105.2838.
\bibitem{ref:monojet}Y. Bai, P. J. Fox, and R. Harnik, J. High Energy Phys. 12, 048 (2010).
\bibitem{ref:IVDM} J.~L.~Feng, J.~Kumar, D.~Marfatia, and D.~Sanford, Phys. Lett. B, {\bf 703}, 2 (2011).
\bibitem{ref:IVDM2} B.~Batell, J.~Pradler, and M.~Spannowsky, arXiv:1105.1781.
\bibitem{ref:IVDM3} J.~Kile and A.~Soni, Phys. Rev. D {\bf 84} 035016, (2011).
\bibitem{ref:IVDM4} P.~Agrawal, S.~Blanchet, Z.~Chacko, and C.~Kilic, arXiv:1109.3516.
\bibitem{ref:IVDM5} S.~Chen and Y.~Zhang, arXiv:1106.4044.
\bibitem{ref:sper3}  T.~Aaltonen {\it et al.}  (CDF Collaboration),  Phys. Rev.  D {\bf 81}, 072003 (2010).
\bibitem{ref:sper1}  T.~Aaltonen {\it et al.}  (CDF Collaboration), Phys. Rev. Lett.  {\bf 106}, 191801 (2011).
\bibitem{ref:sper2}  T.~Aaltonen {\it et al.}  (CDF Collaboration), Phys. Rev. Lett. {\bf 107}, 191803 (2011).
\bibitem{Aad:2011wc} 
 G.~Aad {\it et al.}  (ATLAS Collaboration),
arXiv:1109.4725.
\bibitem{ref:met} Missing transverse energy, $\met$ , is defined as the magnitude of the vector $\vmet = - \sum_i E_T^i \vec{n}_i$ where $E_T^i$ is the magnitude of transverse energy contained in each calorimeter tower $i$, and $\vec{n}_i$ is the unit vector from the interaction vertex to the tower in the transverse $(x,y)$ plane.
\bibitem{ref:CDFII} D. Acosta \textsl{et al.} (CDF collaboration), Phys. Rev. D \textbf{71}, 032001 (2005).
\bibitem{ref:L2upgrade} A.~Bhatti {\it et al.}, IEEE Transactions on Nuclear Science {\bf 56}, 3 (2009).
\bibitem{ref:jetclu} F.~Abe, {\it et al.} (CDF collaboration), Phys Rev. D {\bf 45}, 001448 (1992).
\bibitem{ref:pseudo} CDF uses a cylindrical coordinate system with the z axis along the proton beam axis. Pseudorapidity is $\eta = − \mathrm{ln}(\mathrm{tan}( \theta /2))$, where $\theta$ is the polar angle relative to the proton beam direction, and $\phi$ is the azimuthal angle while $p_T=|p|\mathrm{sin} \theta$, $E_T=E \mathrm{sin} \theta$.
\bibitem{ref:jetcorr} A.~Bhatti {\it et al.}, Nucl. Instrum. Methods A {\bf 566}, 375 (2006).
\bibitem{ref:btag} D.~Acosta {\it et al.} (CDF Collaboration), Phys. Rev. D {\bf 71}, 052003 (2005).
\bibitem{ref:masst}  Tevatron Electroweak Working Group for the CDF and D0 Collaborations,
 arXiv:1107.5255.
\bibitem{ref:topmassreview} A.~B.~Galtieri, F.~Margaroli, and I.~Volobuev,  arXiv:1109.2163.
\bibitem{ref:madgraph}
 J.~Alwall, M.~Herquet, F.~Maltoni, O.~Mattelaer, and T.~Stelzer,  J. High Energy Phys. 06, 128 (2011).
\bibitem{ref:pythia} T.~Sjostrand, S.~Mrenna, and P.~Skands, J. High Energy Phys. 05, 026 (2006).
\bibitem{ref:tagrate} T.~Aaltonen {\it et al.} (CDF Collaboration), Phys. Rev. Lett. {\bf 104}, 141801 (2010).
\bibitem{ref:XsecDIB1} J.~M.~Campbell and R.~K.~Ellis, Phys. Rev. D {\bf 60}, 113006 (1999).
\bibitem{ref:XsecDIB2} J.~M.~Campbell and R.~K.~Ellis, Phys. Rev. D {\bf 62}, 114012 (2000).
\bibitem{ref:Xsectt1} U.~Langenfeld, S.~Moch, and P.~Uwer, Phys. Rev. D {\bf 80}, 054009 (2009).
\bibitem{ref:alpgen} M.~Mangano {\it et al.} J. High Energy Phys. 07, 001 (2003).
\bibitem{ref:XsecST1} B.~W.~Harris {\it et al.}, Phys. Rev. D {\bf 66}, 054024 (2002).
\bibitem{ref:XsecST2} Z.~Sullivan, Phys. Rev. D {\bf 70}, 114012 (2004).
\bibitem{ref:Lumi} D.~Acosta {\it et al.}, Nucl. Instrum. Methods A {\bf 494}, 57 (2002).
\bibitem{ref:tom-CL} T. Aaltonen {\it et al.} (CDF Collaboration), Phys. Rev. D {\bf 82}, 112005 (2010).
\end{thebibliography}
\end{document}